\documentstyle[12pt]{article}
\begin{document}
\begin{center}
{\large\bf
ONE-LOOP VACUUM AMPLITUDE FOR D-BRANES IN CONSTANT ELECTROMAGNETIC FIELD\\}
\vspace{1cm}
A.A. Bytsenko\\
{\it State Technical University, St. Petersburg, 195252, Russia}\\
S.D. Odintsov\\
{\it Institute of Theoretical Physics, Leipzig University,
Augustusplatz 10/11, D-04109 Leipzig, Germany\\
and\\
Tomsk Pedagogical University, 634041 Tomsk, Russia}\\
L.N. Granda\\
{\it Departamento de Fisica, Universidad del Valle\\
A.A. 25360, Cali, Colombia}\\
\vspace{1.5cm}
{\bf Abstract}
\end{center}
Following Polchinski's approach we calculate the one-loop vacuum amplitude for two parallel D-branes connected by open bosonic (neutral or charged) string in a constant uniform electromagnetic (EM) field. For neutral string, external EM field contibution 
appears as multiplier (Born-Infeld type action) of vacuum amplitude without external EM field. Hence, it gives the alternative way to see the inducing of Born-Infeld type action for description of D-branes. For charged string the situation is more 
complicated, it may indicate the necessity to modify the induced D-branes action in this case.\\

\vspace*{2cm}

The importance of non-perturbative relationships between different string theories may require the existence of some extended objects with very specific propierties \cite{wit,hul,sen}. It could be that D-branes [5-7] (see [6] for a review) may play the 
role of such objects. Those objects represent some soliton configurations in superstring theory. They are definitely relevant to the understanding of duality symmetries. From another point, there are some indications [4] that D-branes may also serve for the
 description of black-branes.\\
In refs. [5,6] the vacuum amplitude for two parallel D-branes connected by open bosonic string has been calculated.That gave the way to define the D-brane tension.

Having in mind the possible applications to the explicit construction of D-brane actions, it could be of interest to study the vacuum amplitude for D-branes in external fields. In the present letter we find the vacuum amplitude for parallel D-branes in the 
constant EM field.

We will start from the theory of critical open bosonic string in the external magnetic field
\par$$
S=S_0+S_1\mbox{,}
$$
$$
S_0=-\frac{1}{4\pi\alpha'}\int d^2\sigma\partial_a X_{\mu}\partial^a X^{\mu}
$$
$$
S_1=-\frac{g_0}{2\pi\alpha'}\int d\tau A_{\mu}(x)\delta(\sigma)\dot{X^{\mu}}-
\frac{g_{\pi}}{2\pi\alpha'}\int d\tau A_{\mu}(x)\delta(\sigma-\pi)\dot{X^{\mu}}
\eqno{(1)}$$
where $g_0$ and $g_{\pi}$ are the charges at the ends of the string
(note that for $g_0+g_{\pi}=0$ we have neutral string).

The external magnetic field will be choosen as the following
\par$$
F_{\mu\nu}=\bordermatrix{
                         & & & & &    \cr
                         &0& & & &    \cr
                         & &0&-f_1& & \cr
                         & &f_1&0& &  \cr
                         & & & & ..&    \cr
                         & & & & ..&    \cr}
\eqno{(2)}$$
where $f_i$ are real. The quantization of such model has been discussed in all detail in refs.[11]. Using the results of this quantization, the free energy and Hagedorn temperature for non-critical strings and for critical strings (in magnetic field and 
at non-zero temperature) have been found respectively in refs.[12] and [13].

The mass operator for charged string maybe written as the following [11].
\par$$
\alpha' M^2=\sum_{i=1}^{24-2d}\sum_{n=1}^{\infty}nC_n^{i+}C_n^i+\sum_{i=1}^d\left[\epsilon_ib_0^{i+}b_0^i+\frac{1}{2}\epsilon_i(1-\epsilon_i)+
\right.
$$
$$
\left.\sum_{n=1}^{\infty}(n-\epsilon_i)(a_n^{i+}a_n^i+b_n^{i+}b_n^i)\right]-1
\eqno{(3)}$$
where $tan\gamma_{0i}=g_0f_i$, $tan\gamma_{\pi i}=-g_{\pi}f_i, \epsilon_i=\gamma_{0i}+\gamma_{\pi i}$, d is the number of non-zero $f_i$ in the tensor $F_{\mu\nu}$ (2).\\
Further details of (3) may be found in refs. [11].

At the next step we will discuss the one-loop vacuum amplitude for two parallel D-branes related by open charged string in the magnetic field. As usually [5-7] the Dirichlet boundary conditions for string lead to the modification of mass spectrum [5,6] 
like follows
\par$$
\tilde{M}^2=M^2+\frac{Y\cdot Y}{4\pi^2\alpha'}
\eqno{(4)}$$
where $Y_m=X_1^m-X_2^m$ is the separation of the D-branes, and $M^2$ is given by (3). Hence, we calculate the one-loop vacuum amplitude for D-branes in an external magnetic field.

Following the ref. [6] we may write the one-loop vacuum amplitude as
\par$$
A=\int_0^{\infty} \frac{dt}{2t}\sum_k\exp[-2\pi\alpha't(k^2+\tilde{M}^2)]=
$$
$$
V_{p+1}\int_0^{\infty}\frac{dt}{t}(8\pi^2\alpha't)^{-\frac{p+1}{2}}\exp(-\frac{Y\cdot Y}{2\pi\alpha'})Tr\exp\left\{-2\pi t\times\right.
$$
$$
\left.\left[\sum_{i=1}^{24-2d}\sum_{n=1}^{\infty}nC_n^{i+}C_n^i+\sum_{i=1}^d\left[\epsilon_ib_0^{i+}b_0^i+\frac{1}{2}\epsilon_i(1-\epsilon_i)+\right.\right.\right.
$$
$$
\left.\left.\left.\sum_{n=1}^{\infty}(n-\epsilon_i)(a_n^{i+}a_n^i+b_n^{i+}b_n^i)\right]-1\right]\right\}
\eqno{(5)}$$
Note that in the absence of magnetic field, i.e. $d=0$ we are back to Eq. (2.23) of ref. [6].

Calculating the traces over the entire Fock space we get:
\par$$
A=V_{p+1}\int_0^{\infty}\frac{dt}{t}\left(8\pi^2\alpha't\right)^{-\frac{p+1}{2}}\exp\left(-\frac{Y\cdot Yt}{2\pi\alpha'}+\frac{\pi dt}{6}\right)\eta^{2d-24}(it)\times
$$
$$
\prod_{i=1}^{d}\frac{e^{-\pi t\epsilon_i(1-\epsilon_i)}}{1-e^{-2\pi t\epsilon_i}}\prod_{i=1}^d\prod_{n=1}^{\infty}\left(1-e^{-2\pi t(n-\epsilon_i)}\right)^{-2}
\eqno{(6)}$$
Here 
\par$$
\eta(\tau)=e^{\frac{i\pi\tau}{12}}\prod_{n=1}^{\infty}\left(1-e^{2\pi in\tau}\right)
\eqno{}$$
The asymptotics at $t->0$ are given in ref. [12]
\par$$
\prod_{i=1}^d\prod_{n=1}^{\infty}\left(1-e^{-2\pi t(n-\epsilon_i)}\right)^{-2}=\prod_{n=1}^{\infty}\left(1-e^{-2\pi tn}\right)^{-2d}\times
$$
$$
\left(2\pi t\right)^{-2\sum_{i=1}^d\epsilon_i}\exp\left((2\gamma +\pi t)\sum_{i=1}^d\epsilon_i+0(t^3)\right)
\eqno{(7)}$$
with $\gamma$ being the Euler-Mascheroni constant. One can also use the modular transformation in order to obtain the $t->0$ asymptotics, namely
\par$$
\eta^{2d-24}(it)=t^{12-d}\exp\left(\frac{2\pi}{t}(1-\frac{d}{12})\right)\prod_{n=1}^{\infty}(1-e^{-2\pi n/t})^{2d-24}\mbox{,}
$$
$$
\prod_{n=1}^{\infty}(1-e^{-2\pi nt})^{-2d}=t^d\exp\left(-\frac{\pi td}{6}+\frac{\pi d}{6t}\right)\prod_{n=1}^{\infty}(1-e^{-\frac{2\pi n}{t}})^{-2d}
\eqno{(8)}$$
Finally, we obtain
\par$$
A=V_{p+1}\int_0^{\infty}\frac{dt}{t}(8\pi^2\alpha't)^{-\frac{p+1}{2}}\exp\left(-\frac{Y\cdot Yt}{2\pi\alpha'}+(2\gamma+\pi t)\sum_{i=1}^d\epsilon^2_i\right)\times
$$
$$
t^{12}\left(2\pi t\right)^{-d-2\sum_{i=1}^d\epsilon_i}\prod_{i=1}^d\epsilon_i^{-1}\left(e^{2\pi/t}+24+o(e^{-2\pi/t})\right)
\eqno{(9)}$$
The massless pole from the second term in (9) is
\par$$
A\sim V_{p+1}\cdot 24\cdot 2^{-12-d-2\sum_{i=1}^d\epsilon_i}\left(4\pi^2\alpha'\right)
^{11-p-d-2\sum_{i=1}^d\epsilon_i} \pi^{\frac{p-23}{2} -d-2\sum_{i=1}^d\epsilon_i}\times
$$
$$
\prod_{i=1}^d\epsilon_i^{-1}\exp\left(2\gamma\sum_{i=1}^d\epsilon_i\right)\left(Y\cdot Y-2\pi^2\alpha'\sum_{i=1}^d\epsilon_i^2\right)^{\frac{p+1}{2}+d+2\sum_{i=1}^d\epsilon_i-12}
\eqno{(10)}$$
For $d=0$ Eq. (10) coincides with expression (2.27) in ref. [6].
However, the interpretation of the expression (10) in the spirit of refs. [5,6] as scalar Green function (up to some coefficient) is not so evident.\\
In particular, one sees that due to the presence of $2\sum_{i=1}^d\epsilon_i$ in the power of the last multiplier in (10), this last multiplier may be in non-integer power (unlike the situation in refs. [5,6]).

From another point, considering eq. (10) as a kind of static potential for D-branes one sees that for short distance between D-branes the presence of magnetic field may lead to effective change of D-branes separation. It maybe that the analysis of the 
effective massless particles limit of operator (3) may help in better understanding of (10).

Let us turn now to neutral open bosonic string where situation is much simpler. In this case the mass operator is given by the expression (see [8] and second reference of [11])
\par$$
\alpha'M^2=\sum_{i=1}^{24-2d}\sum_{n=1}^{\infty}nC_n^{i+}C_n^i-\alpha'\sum_{j=1}^{2d}\frac{p_jg_j^2f_j}{1+g_j^2f_j^2}+
$$
$$
+\sum_{i=1}^d\sum_{n=1}^{\infty}n\left(a_n^{i+}a_n^i+b_n^{i+}b_n^i\right)-1
\eqno{(11)}$$
where $g_0+g_{\pi}=0$, $g_i$ are the charges of the interaction with 
the external field in the i-th dimension.\\
Putting $g_j=1$ and making the calculation very similar to the one given in ref. [8] (see also [11] for non-zero temperature case) we find:
\par$$
A=V_{p+1}\prod_{i=1}^{2d}(1+f_i^2)\int_0^{\infty}\frac{dt}{t}\left(8\pi^2\alpha't\right)^{-\frac{p+1}{2}}\exp(-\frac{Y.Yt}{2\pi\alpha'})\times
$$
$$
\times t^{12}\left(\exp(\frac{2\pi}{t})+24+...\right)
\eqno{(12)}$$
where the mass operator (11) with $Y.Y$ term (like in (4)) has been used. Because
\par$$
\prod_{i=1}^{2d}(1+f_i^2)=det(1+F)
\eqno{(13)}$$
is just the Born-Infeld type action in open string theory [9], the vacuum amplitude (12) is proportional to the result of refs. [5,6] (up to the factor (13)). Hence, the interpretation of refs. [5,6] is valid, and the tension $T_p$ is given by the
 same expression (2.29) of ref. [6].

Note that in ref. [10] and recent preprints [14-16] it has been shown (using other arguments) that D-brane actions may be described by Born-Infeld type effective action (here, duality transformations may be found in a similar way as in non-linear 
electrodynamics [17]).

Hence, we gave the alternative way to show the inducing of Born-Infeld type action for description of D-branes from the study of vacuum amplitude of neutral open bosonic string in the external magnetic field. The fact that for charged string the situation
 is more complicated may indicate the necessity of modifying of induced D-branes action in this case.

As the last example, we will consider an open non-critical neutral string in the constant uniform EM field in D-dimensional space. The mass operator is given by [11,12]
\par$$
\alpha^\prime M^2=-\alpha^\prime\sum\limits_{i=1}^{(D-1)/2}
\frac{e^2+h_i^2}{1+h_i^2}[(p_{2i}^2)^2+(p_{2i+1}^2)^2]+\frac{d_0^2}{2}+ $$
$$+(1-e^2)\left[\sum\limits_{n=0}^\infty\,\sum\limits_{j=1}^{(D-1)/2}
\,n(a^{2j+}_n a^{2j}_n+a_n^{2j+1+}a^{2j+1}_n)-\frac{D-1}{24}\right]
+\frac{Y\cdot Y}{4\pi^2\alpha^\prime}
\eqno{(14)}$$
where periodic length L of the Weyl mode [19] appears in this expression (see also [12]).

The calculation of the vacuum amplitude may be done in the same way as in previous case with the following result:
\par$$
A=\int\limits_0^\infty\frac{dt}{t}\sum\limits_k
e^{-2\pi\alpha^\prime t(k^2+M^2)}=V_{p+1}(2\pi)^{-1/2}
(8\pi^2\alpha^\prime)^{-\frac{p+1}{2}}[\pi(1-e^2)]^{-\frac{D+1}{2}}\times$$
$$\times{\rm det}(1+F_{\mu\nu})\int\limits_0^\infty\,dt\cdot t^{-2-p/2}
\exp\left(-\frac{Y\cdot Y\cdot t}{2\pi\alpha^\prime}\right)
\eta^{1-D}[i\pi t(1-e^2)]\theta_3(0,e^{-L^2/2\pi t})
\eqno{(15)}$$
Hence, again as in previous case the Born-Infeld type action factor appears in front of the integral. The whole expression after exclusion of the Born-Infeld type action gives the generalization of Polchinski's result for non-critical open string (for D=
25 it coincides with the result of ref. [6] or Eq. (12).)

For $D=1$ the expression (15) my be rewritten in a simple form:
\par$$
A=V_{p+1}\frac{(8\pi^2\alpha^\prime)^{-\frac{p+1}{2}}}
{\sqrt{\pi}L(1-e^2)}{\rm det}(1+F_{\mu\nu})\times$$
$$\times\int\limits_0^\infty\,dt\cdot t^{-\frac{p+3}{2}}
\sum\limits_{n=-\infty}^\infty\exp\left[-t\left(
\frac{Y\cdot Y}{2\pi\alpha^\prime}+\frac{2\pi^3n^2}{L^2}
\right)\right]
\eqno{(16)}$$
Note that in the absence of D-branes separation this integral may be expressed in elementary functions [20].

Thus, for open string in a constant electromagnetic field the inducing of Born-Infeld type action for description of D-branes from the study of the correspondent vacuum amplitude is again shown. Note that generalization to supersymmetric case (where again
 Born-Infeld type action is induced) is straitforward. Recently, brane-antibrane forces for such objects have been discussed in [18].\\

It would be of interest to discuss the behaviour of D-branes in gravitational field (of the De Sitter type for example), what can be done in a similar way.

\vspace*{1cm}

\bf Acknowledgments\rm \\
SDO would like to thank R. Iengo for useful remark. This work has been supported by GRASENAS 95-0-6.4-1

\end{document}